# Room-temperature continuous-wave operation of organometal halide perovskite lasers

Zhitong Li[1†], Jiyoung Moon[2†], Abouzar Gharajeh[1†], Ross Haroldson[3], Roberta Hawkins[2], Walter Hu[1,4], Anvar Zakhidov[3,5], Qing Gu[1]*

[1]Department of Electrical and Computer Engineering, [2]Department of Material Science and Engineering, and [3]Department of Physics, The University of Texas at Dallas, Richardson, TX 75080
[4]ASIC and System State-key Lab, Institute of Microelectronics, Fudan University, Shanghai 200433, China
[5]Department of Nanophotonics and Metamaterials, ITMO University, St. Petersburg, Moscow, Russia
*Corresponding author. Email: qing.gu@utdallas.edu

†These authors contributed equally to this work

Solution-processed organic-inorganic lead halide perovskites have recently emerged as promising gain media for tunable semiconductor lasers. However, optically pumped continuous-wave lasing at room temperature – a prerequisite for a laser diode – has not been realized so far. Here, we report lasing action in a surface emitting distributed feedback methylammonium lead iodide (MAPbI$_3$) perovskite laser on silicon substrate, at room temperature under continuous-wave optical pumping. This outstanding performance is achieved because of the ultra-low lasing threshold of 13 W/cm$^2$, which is enabled by thermal nanoimprint lithography that directly patterns perovskite into a high Q cavity with large mode confinement, while at the same time improves perovskite's emission characteristics. Our results represent a major step toward electrically pumped lasing in organic and thin-film materials, as well as the insertion of perovskite lasers into photonic integrated circuits for applications in optical computing, sensing and on-chip quantum information.

Since the advent of silicon (Si) photonics, the field of photonic integrated circuit (IC) has progressed significantly over the last few decades [1]. While many photonic components have the potential to be inserted into future electronic-photonic ICs on Si, a critical component – an efficient chip-scale laser on Si – has not been realized because of Si's indirect bandgap. Although III-V/Si lasers formed via wafer bonding of III-V onto Si substrate have been the main candidate, the low yield and high manufacturing cost restrict their further development as light sources for electronic-photonic ICs [2]. An alternative gain medium that is Si compatible is solution-processed organic semiconductors. Although organic lasers cannot compete with inorganic III-V lasers in many performance metrics, they do offer several advantages such as easy wavelength tunability, low-cost processing and mechanical flexibility. In fact, the Si-organic hybrid approach has already been applied in Si photonics for the construction of waveguides and modulators [3]. However, room temperature continuous-wave (CW) lasing has not been achieved in optically pumped organic semiconductor lasers, and their future advancement into electrically pumped lasers remains an elusive challenge [4]. To this end, solution-processed hybrid organic-inorganic perovskites have emerged as the gain material candidate that may satisfy the requirements for on-chip optical sources, namely, Si compatibility, electronic addressability and cost-effective fabrication. In addition to easy wavelength tunability [5], strong optical absorption, high carrier mobility [6] and long-range carrier diffusion length [7], the balanced ambipolar charge transport characteristics of perovskites may hold the key to realizing electrically pumped solution-processed lasers [8,9].

Despite the promising material properties, the lack of a room temperature CW pumped perovskite laser has become the bottleneck of its possible application in photonic ICs. Since the inception of perovskite laser in 2014 [10], a variety of optically pumped lasers have been demonstrated with ultrashort pump pulses (pulse durations of less than 5 ns) and/or at cryogenic temperatures, mostly on low-index substrate such as glass or mica [11–31]. In these works, most use the hybrid methylammonium lead iodide ($MAPbI_3$) as the gain material; meanwhile, lasers using other hybrid perovskites (such as $FAPbI_3$ and $MAPbBr_3$) as well as all-inorganic perovskites are also emerging [18–21,23,31]. Recently, the first CW lasing action was reported at 100 K in a $MAPbI_3$ distributed feedback (DFB) cavity [17]. This was achieved by exploiting a mixture of tetragonal and orthorhombic phases of $MAPbI_3$, which formed a material structure analogous to inorganic semiconductor quantum well, below the tetragonal-to-orthorhombic phase transition temperature of 160 K. This work marked a major step toward realizing a room temperature CW optically pumped perovskite laser. Nevertheless, when CW operation was attempted at temperatures above 160 K, lasing was found to cease within tens of nanoseconds following pump turn-on. It was suspected that the absence of lasing behavior at higher temperature is a result of photo-induced structural change in perovskites that in turn reduces the material gain on a sub-microsecond time scale, which is to be expected under the high peak pump density exceeding 5 $kW/cm^2$ [11]. In fact, it was speculated that the pump power required for $MAPbI_3$ to lase under CW pumping would be ~14 $kW/cm^2$ [15]. In this context, further breakthroughs are required for the realization of optically pumped perovskite lasers operating in the CW regime at room temperature, and ultimately, electrically pumped perovskite laser diodes.

Here, we report the first room temperature CW lasing action from perovskites, in a $MAPbI_3$ DFB resonator on the Si platform. With pump density of a mere 13 $W/cm^2$, which is three orders of magnitude lower than the lasing threshold values reported in existing literature, lasing is achieved and sustained in ambient air environment. The two prominent features of our laser that give rise to the ultra-low threshold are: i) direct patterning of perovskite to form a high Q-factor cavity with large mode-gain overlap, and ii) improvement of material's emission characteristics, both of which are accomplished by thermal nanoimprint lithography (NIL). These results suggest a new strategy to design and manufacture ultra-low threshold perovskite lasers on the Si photonics platform, and marks a major advancement in the field of perovskite lasers. This work also opens new prospects toward the development of a solution-based electrically pumped laser diode.

**Results**

Our surface-emitting perovskite laser is constructed by directly patterning a $MAPbI_3$ thin film by NIL, on silicon dioxide thermally grown on a Si substrate. Fig. 1a sketches the principle of NIL and the resulting perovskite cavity. It is important to note that the commonly used e-beam lithography cannot be applied to perovskites due to material instability under moisture and solubility in many solvents. Instead, cavities are typically formed by solution-phase growth or spin-coating methods, both of which lack dimension control, repeatability and uniformity. On the other hand, although the solvent-free NIL technique has been widely used in patterning organic materials, because perovskites are a hard, ionic-based material without a glass transition behavior, NIL was believed to be not applicable to perovskites until recently [32,33]. By employing NIL, we therefore not only directly pattern perovskites with pre-defined cavity geometry, but also offer excellent dimension control and repeatability. At elevated temperature and pressure, the e-beam lithography fabricated stamp is pressed against spin-coated $MAPbI_3$ (Fig. 1a), thus forming a DFB cavity with $MAPbI_3$. The dimensions of the $MAPbI_3$ cavity is the inverse replica of the stamp. Fig. 1b shows a perspective view Scanning Electron Microscope (SEM) image of a representative device, highlighting the clear cavity definition and the low surface roughness of $MAPbI_3$ enabled by NIL. The details of $MAPbI_3$ thin film preparation, stamp fabrication and NIL process are provided in Methods.

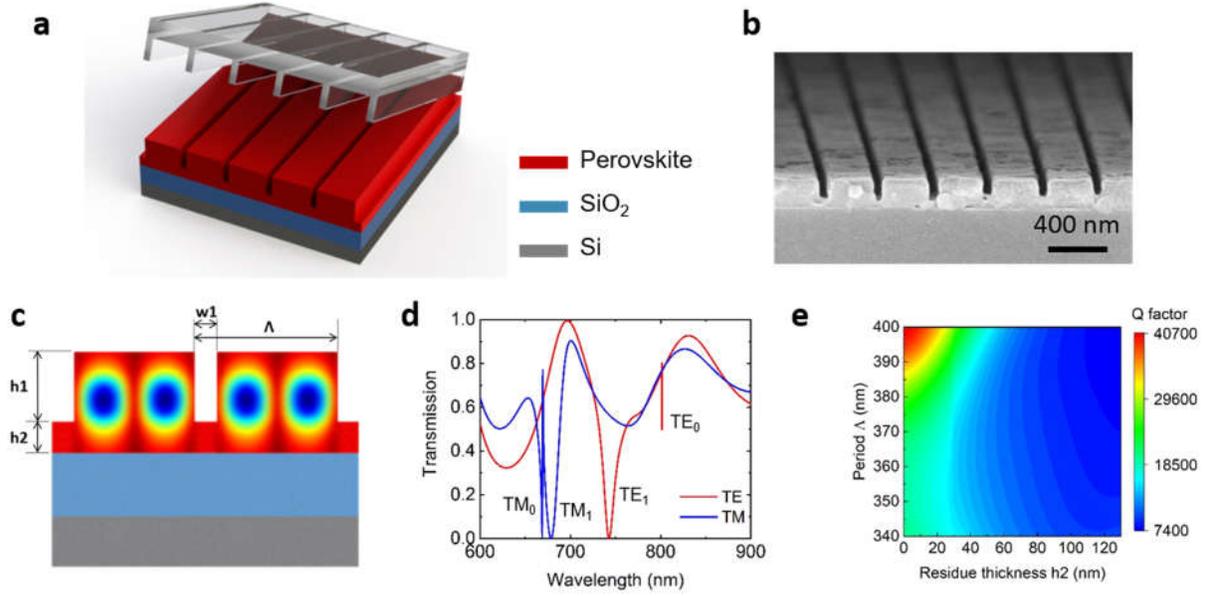

**Fig. 1** Fabrication scheme and cavity design of surface-emitting perovskite DFB lasers. **a** Schematic illustration of the nanoimprint process. **b** Perspective view SEM image of the imprinted MAPbI$_3$ DFB resonator. **c** Side-view of the DFB cavity with the simulated TE$_0$ mode profile superimposed. Blue color represents field antinode. **d** Simulated transmission spectrum for the device geometry depicted in **Fig. 1**b where air trench width w1 = 50 nm, grating period Λ = 380 nm, grating height h1 = 145 nm and perovskite residue thickness h2 = 30 nm. **e** Simulated dependence of the cavity Q factor on the grating period and the air trench width, for the TE$_0$ mode.

The threshold gain g$_{th}$ of a lasing mode is defined as $g_{th} = n\omega_0/c\Gamma Q$, where ω$_0$ is the modal frequency, n is the refractive index of the gain medium, c is the speed of light in vacuum. Γ is the mode-gain overlap, related to how well the cavity confines the mode to the gain region, while Q is the mode's quality factor, related to the amount of radiation and material absorption loss the mode experiences. Efforts to reduce lasing threshold in perovskite lasers focus on either increasing Q (such as microdisk and photonic crystal based designs [10,12,34]) or increasing Γ (such as vertical cavity surface emitting laser based designs [21,35]). With the capability to directly pattern perovskites, we aim to increase Γ and Q simultaneously. To achieve low threshold lasing with emission in the direction normal to the substrate plane, we employ a second order DFB design. Using the finite element method (FEM) analysis in the wave optics module of COMSOL Multiphysics, we perform a comprehensive parametric sweep to determine the optimal cavity geometry that supports a low threshold gain mode, with a resonant wavelength within the MAPbI$_3$ photoluminescence (PL) range. A typical fabricated DFB cavity with the optimized design is shown in Fig. 1b, whose geometrical parameters are chosen to support the fundamental transverse electric (TE$_0$) mode that satisfies the second order Bragg diffraction condition $m\lambda_{Bragg} = 2n_{eff}\Lambda$, where m = 2 corresponds to the second order diffraction, λ$_{Bragg}$ is the Bragg wavelength, n$_{eff}$ is the effective index of the mode, and Λ is the grating period. The transmission spectrum in Fig. 1d shows that TE$_0$ mode has a resonant wavelength of 802 nm and a Q factor of ~2×10$^4$, and a mode-gain overlap of 93% (illustrated by the projection of the mode profile onto the side-view of the DFB cavity in Fig. 1c). In addition to the targeted TE$_0$ mode with the highest Q-factor, as suggested by its narrowest linewidth among all modes, Fig. 1d also depicts the simulated transmission spectrum of other modes that reside in the spectral window of gain, at near-normal incidence (incident angle of 0.5° to ensure finite Q-factor at high-symmetry points). Lastly, we present in Fig. 1e how one may tune the Q factor by changing the grating period Λ or the perovskite residue thickness h2, the latter of which is possible only with NIL.

To further understand NIL's impact on our system, we analyze MAPbI$_3$'s property before and after nanoimprinting. We prepare two samples with 150 nm thick MAPbI$_3$ thin film, for which the thickness is chosen to be the same as that used for the construction of the DFB resonator shown in Fig. 1b, and apply NIL on one sample with a flat stamp. Subsequently, we perform X-ray diffraction (XRD), SEM and Atomic Force Microscopy (AFM) on both samples. Fig. 2a and b depict the SEM images of the pristine and the imprinted thin films, respectively. While the pristine polycrystalline MAPbI$_3$ has domain sizes of a few hundred nanometers, the imprinted film shows much larger domain sizes, indicating the assembly of small crystallites into a tiling of larger crystals. The AFM images (inset of Fig. 2a and b) further illustrate the improved surface morphology by NIL: the surface roughness reduces from 22 nm (rms) of the pristine film to 4 nm of the imprinted film. Fig. 2c shows a comparison of the XRD spectra. The increased intensity of the (220) reflection suggests the preferred orientation of MAPbI3 crystals, while the emergence of the PbI$_2$ (001) peak in the imprinted film suggests material degradation, as a result of the elevated temperature and pressure during NIL. Note that the sharp peak at 33° in the NIL sample is known as the basis-forbidden Si (200) peak, which depends on the orientation of the sample and has been well investigated [36]. A trade off therefore exists between the material morphology and the impurity level. The SEM and XRD results suggest that NIL induces the formation of larger and more ordered domains. We also investigate the stability of MAPbI$_3$ by performing XRD on both pristine and nanoimprinted films after 15 days (Supplementary Figure 4b). The decomposition rate of MAPbI$_3$ to PbI$_2$ can be estimated by the ratio of peak intensities between the (001) PbI$_2$ and the (110) MAPbI$_3$ peaks. The imprinted film has a decomposition rate of 0.316 whereas the pristine film has a rate of 0.397. This result implies that the material decomposition rate is slowed down through nanoimprinting. The combined effects of the improved morphology and the increased impurity level on the emissive property of MAPbI$_3$ is presented in Fig. 2d. The imprinted film shows a higher slope efficiency, mainly contributed by the reduced scattering loss due to NIL's reduction of thin film surface roughness.

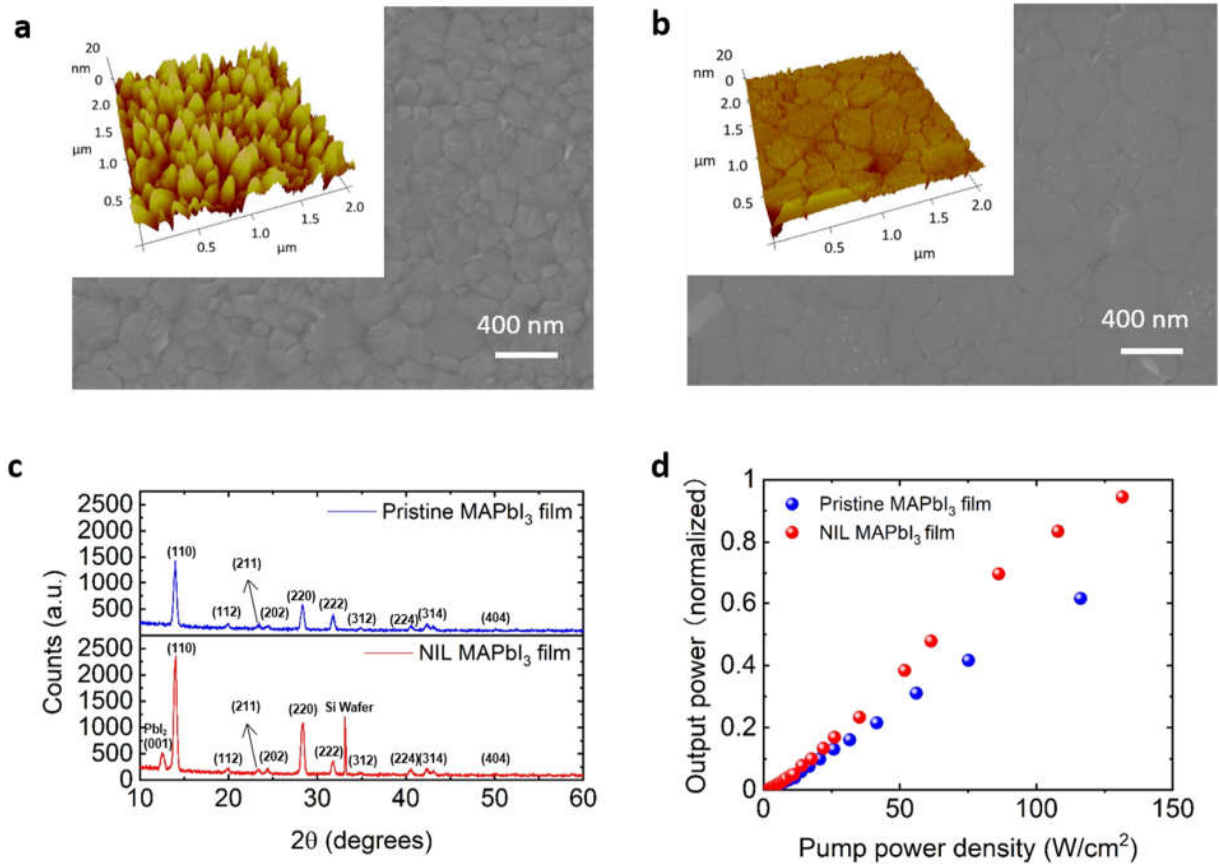

**Fig. 2** Effect of NIL on MAPbI$_3$'s morphology and emissive property. SEM (inset: AFM) images of **a** pristine and **b** imprinted thin film, showing larger domain size and smoother surface of perovskite after NIL. Scale bar: 400 nm. **c** XRD spectra of pristine and imprinted thin films with assignment of the most relevant peaks. **d** Comparison of the photoluminescence output characteristics of pristine and imprinted thin films upon CW pumping at room temperature.

Fig. 3 shows the emission characteristics of our perovskite laser under room temperature CW excitation, in the ambient air environment without any substrate cooling. We project the horizontally polarized 355 nm pump beam onto the sample with a spot size of 20 µm, and record emission's steady-state spectral behavior. Using a cascaded 4-f imaging system in conjunction with a pump filter and a TE-pass polarizer, emission is collected by a spectrograph with a cooled Si detector. Fig. 3a depicts the evolution of the output power as a function of both the pump power and the wavelength, from a very low-intensity broad spectrum at low pump levels to a dominant lasing peak at high pump levels. The lasing peak at 807 nm has a full width at half maximum

(FWHM) of 1.2 nm, although FWHM as narrow as 0.7 nm is observed in samples with similar dimensions (Fig. 3c). This linewidth is on par with the lasing FWHM of reported perovskite lasers, which ranges from 0.1 nm to 2.2 nm, although those works all used pulsed pumping and/or cryogenic temperatures in which narrower linewidths are to be expected [11,21,37,38]. Fig. 3b shows the light-in vs. light-out characteristics (light-light curve) of the emission from $TE_0$ mode, which has a slope change, indicating the onset of lasing, at the external threshold pump density of ~ 13 W/cm$^2$. This strikingly low threshold pump power density, which is a few orders of magnitude lower than existing perovskite lasers, is achieved because of the high Q-factor and the large confinement factor in the directly patterned cavity, in addition to the improved emission characteristics of perovskite after NIL. It is this low lasing threshold that ensures the room temperature operation.

The dependence of $TE_0$ mode's emission wavelength on pump power is presented in Fig. 3d, where an overall blue-shift of 0.38 nm is observed over the two-orders-of-magnitude range of pump levels. Blue-shift of emission with increasing pump power is indicative of perovskite self-heating, which is significant under CW pumping at room temperature. However, compared with previously reported values [11,22,39,40], the 0.38 nm shift is minimal, asserted by the extremely low pump power used to achieve lasing. Note that because of the spectrograph wavelength step size of 0.19 nm used in obtaining the spectra, the evolution of emission wavelength shows a step-like behavior rather than the actual gradual shift. Nonetheless, with poor thermal conductivity of < 1.5 W/mK in $MAPbI_3$ at room temperature [41], large Auger recombination, decreased material gain and material degradation are to be expected. To test the stability of the CW lasing behavior, we record the intensity of the lasing peak over time. As shown in Fig. 3e, a slight rise in intensity over the first 120 seconds is observed, followed by a drop to below the initial value, consistent with the trend observed under CW pumping at cryogenic temperature [17]. After 230 seconds, the intensity drops to less than 75% of the initial value, accompanied by the disappearance of the emission peak. Although the lasing duration is much longer than previous attempts under similar excitation conditions [11], self-heating still has an adverse effect on device performance. Furthermore, because all measurements are performed in ambient atmosphere, in addition to self-heating, material degradation due to moisture and oxygen exposure is to be expected during the duration of the measurements. Note that the data presented in Fig. 3 are from multiple samples with dimensions similar to that shown in Fig. 1b, i.e. with 380 nm grating period, 50 ± 5 nm air trench width, 145 ± 10 nm grating height, and 30 ± 10 nm residue thickness.

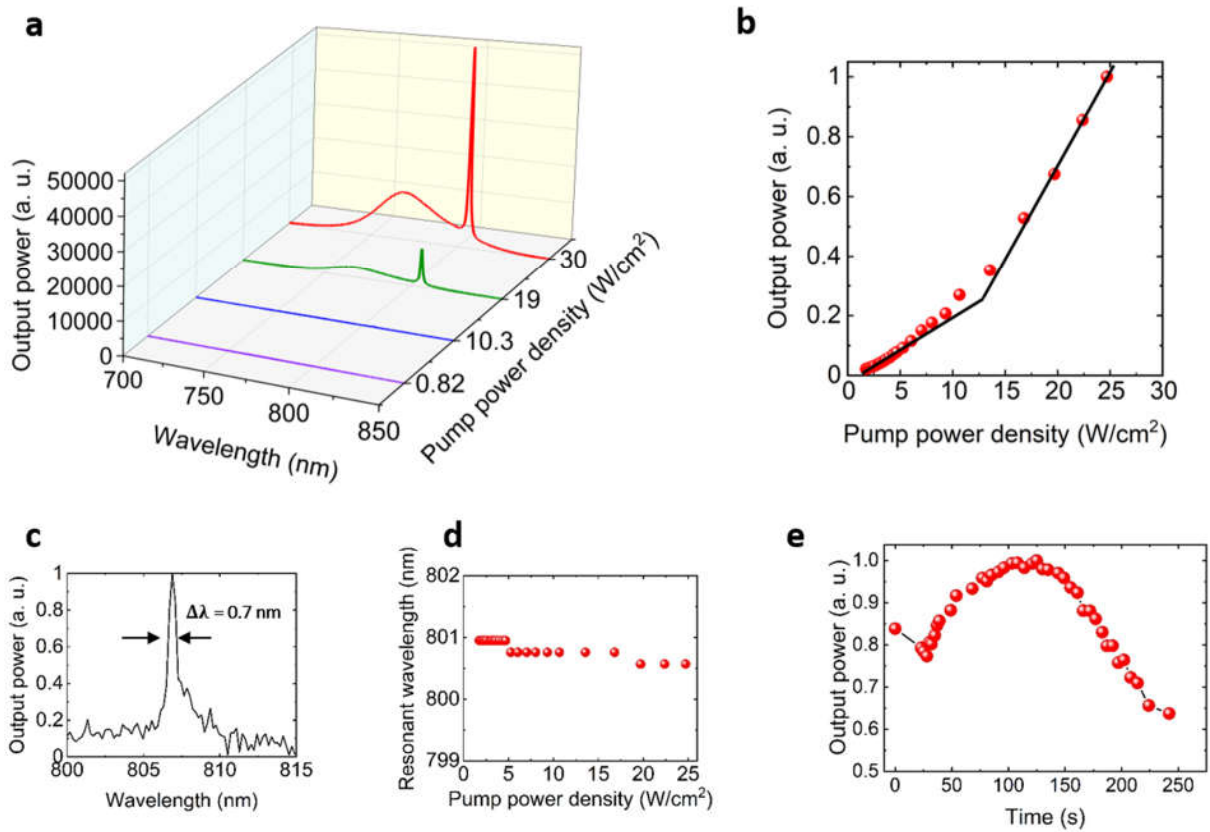

**Fig. 3** Continuous-wave lasing characteristics at room temperature. **a** Output power as a function of the pump power density (light-light curve) around the lasing wavelength. **b** Evolution of the normalized output power as a function of wavelength and pump power density. **c** Zoom-in spectrum showing a 0.7 nm linewidth around 807 nm. **d** Peak emission wavelength as a function of pump power density. **e** Emission intensity at ~15 W/cm$^2$ pump power density over the duration of 250 s, recorded from a single excitation spot.

Although we can indeed precisely control the optical mode characteristics (resonant wavelength and cavity Q factor) of the DFB laser by changing the NIL stamp dimensions such as the air trench width and the grating period, each iteration requires the fabrication of a new stamp via e-beam lithography. Another unique feature of our nanoimprinted DFB laser design is the additional degree of freedom to tune the optical mode characteristics by simply adjusting the perovskite residue thickness ($h_2$), without any additional stamp fabrication step. The experimentally measured TE$_0$ mode resonant wavelength as a function of the residue thickness $h_2$ is presented in Fig. 4a, in which the resonant wavelength red-shifts with increasing residue thickness. Fig. 4a also shows the resonant wavelength's dependence on the grating period Λ. We observe that the residue thickness and the grating period affect the resonant wavelength in similar fashions, namely, red-shifting the resonance with increasing parameter value. Excellent agreement can be seen between the experimental data (Fig. 4a) and the simulated results (Fig. 4b) in terms of the evolution profile, albeit with a 20 nm wavelength shift. The wavelength difference can be accounted for by the ± 10 nm error range in the residue thickness measurement by SEM, as well as the deviation of the MAPbI$_3$ refractive index used in simulation (n ≈ 2.52) from the actual value.

It is important to note that, although the grating period and the perovskite residue thickness can both be used to tune the emission wavelength, they affect laser performance differently. First, the grating period and the residue thickness affect the Q factor in opposite directions, as shown in Fig. 1e for TE$_0$ mode. Second, increasing the residue thickness also increases the number of resonant modes, which in turn encourage mode competition that typically reduces the efficiency of the laser. This situation is illustrated in Fig. 4c and d for TE and TM modes, respectively. Using the cavity geometrical parameters of the device depicted in Fig. 1b, but with various residue thicknesses, Fig. 4c shows the transmission spectrum of TE modes at near-normal incidence (incident angle of 0.5°), where the narrowest linewidth of TE$_0$ mode suggests that it has the highest Q factor among all modes. As the residue thickness increases, in addition to the increase of Q factor for some modes, the resonant wavelength red-shifts, a combination of which result in the inclusion of more cavity modes with appreciable Q factor in the spectral window of MAPbI$_3$ gain. Fig. 4d presents a similar situation for TM modes. Combining the effect of the residue thickness on wavelength, Q factor and the number of high-Q modes, resonance tuning by perovskite residue thickness should therefore be used with caution. Thin residue thickness is generally preferred. Luckily, because resonant wavelength is adjusted at an average rate of 0.5 nm per 1 nm change in residue thickness, a tuning range of a few tens of nanometers can be achieved while ensuring single TE$_0$ mode operation, as long as the residue thickness is kept below 70 nm. Because residue-free NIL is more challenging, we design our DFB lasers with 30 nm perovskite residue thickness.

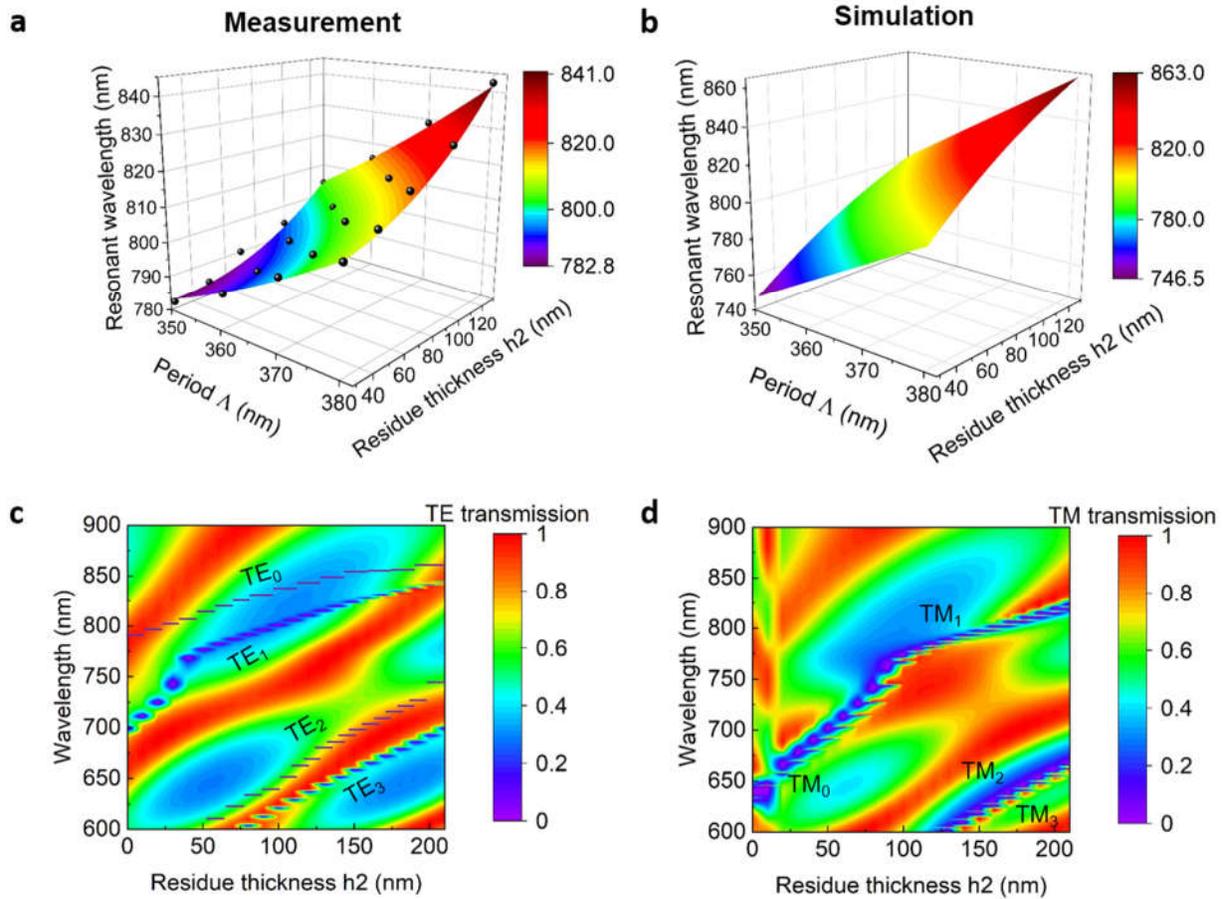

**Fig. 4** Tunability of emission wavelength via grating period and perovskite residue thickness. **a** Measured and **b** simulated resonant wavelength of $TE_0$ mode as a function of grating period and perovskite residue thickness. In **a**, the black dots are measurement points and the surface plot is the fitting of the measured data. **c-d** Simulated transmission spectra for **c** TE and **d** TM modes as a function of perovskite residue thickness, with air trench width w1 = 50 nm, grating period Λ = 380 nm and grating height h1 = 145 nm. Dashed magenta lines are added in **c** for visual aid to illustrate the high-Q mode with vanishing linewidth.

## Discussion

By utilizing a second order DFB resonator directly formed with $MAPbI_3$, we have demonstrated CW lasing in perovskite laser at room temperature, with ultra-low pump power density of 13 W/cm$^2$ at lasing threshold. In addition to increasing the cavity Q factor, the mode-gain overlap and perovskite's emissive properties, the NIL technique can also be utilized to tune the laser's emission wavelength by adjusting the perovskite residue thickness that results from nanoimprinting. In so doing, only one NIL stamp is needed for multiple resonator geometries, making our approach a cost-effective and manufacturing friendly one for realizing perovskite lasers and other perovskite optoelectronic devices.

Despite the successful lasing behavior in ambient atmosphere, laser-induced thermal degradation, as well as moisture and oxygen exposure induced degradation of MAPbI$_3$ gain medium prevent sustained lasing over extended period of time. Further work to significantly enhance the room temperature CW operational stability of perovskite lasers is therefore crucial. To this end, the intrinsic material stability can be improved via anion and/or cation mixing, or by utilizing the more stable low dimensional perovskite as the gain medium [42–44]. In parallel, devices' humidity tolerance can be improved with perovskite encapsulation, by incorporating, for example, alternating atomically thin layers of metal oxide and hydrophobic film into the cavity design. To minimize self-heating, high thermal conductivity dielectric such as aluminum oxide (Al$_2$O$_3$) or aluminum nitride (AlN) can be used in place of SiO$_2$ in the cavity design to assist heat dissipation through substrate. The integration of additional heat dissipation component into the laser can also be considered. Combined with these improvements, a stable and low-threshold perovskite laser opens the door to realize the first solution-processed laser diode, for applications in optical and data communication, sensing and display technologies.

## Methods

### Perovskite thin film preparation

The MAPbI$_3$ solution is prepared by dissolving a 1:1 molar ratio of PbI$_2$ and CH$_3$NH$_3$I in a 1:1 volume ratio of γ-butyrolactone (GBL):dimethylformamide (DMF) solvent mixture in a N$_2$ glovebox, resulting in a concentration of 1.5M. Dymethyl sulfoxide (DMSO) is added into the solution such that a 1:1 molar ratio of PbI$_2$:DMSO is achieved. The solution is subsequently heated for 24 hours at 60 °C and then diluted with 1:1 GBL:DMF to achieve the targeted thickness. In the meantime, Si substrates covered with a thermally grown SiO$_2$ layer of 1 μm thickness are cleaned by ultrasonication in acetone and treated with UV-Ozone for 15 minutes. Next, a two-step spin-coating process, at 1000 rpm for 23 s followed by 4000 rpm for 30 s, is performed. 300 microliters of anhydrous toluene is dropped on the film 12 s into the second spin-coating step. Lastly, the sample is annealed on a hot plate at 100 °C for 10 minutes, during which solvents are evaporated, and a dense and uniform MAPbI$_3$ film is formed with thicknesses in the range of 147 nm – 247 nm (for perovskite residue range of 30 nm – 130 nm after NIL, per cavity design).

### Stamp fabrication and thermal nanoimprint lithography (NIL) process

The NIL stamp is designed to have trench width of 50 nm, trench depth of 140 nm, and period in the range of 350 nm to 380 nm with 10 nm increment. To fabricate the SiO$_2$ stamp, a 500 nm thick SiO$_2$ layer is first thermally grown on Si substrate, followed by the e-beam evaporation of a 80 nm thick layer of Chromium (Cr) as the etch hard mask. 35 nm thick 2% hydrogen silsesquioxane (HSQ) e-beam resist (Dow Corning® XR-1541) and 30 kV e-beam writing condition (Raith 150TWO e-beam lithography system) are chosen to produce ~40 nm wide gratings with good grating definition. Prior to spin-coating, a filter with pore size of 0.45 µm is used to filter out large particles in HSQ. Subsequently, HSQ is spin-coated in a two-step process (step 1: 500 rpm for 5 s, step 2: 3000 rpm for 60 s) on the Cr/SiO$_2$/Si stack of 3×3 cm$^2$ in size, and pre-baked at 90 °C for 5 mins. The writing condition is optimized by repeating the process of focusing, aperture alignment and stigmatism correction. Because the sample surface is likely to be tilted with respect to the electron beam, leveling of the sample surface is performed to obtain uniform electron beam condition on all patterns. After e-beam writing, the HSQ is developed in tetramethylammonium hydroxide (TMAH) 25% for 1 min at 35 °C. The exposed HSQ serves as a mask for the subsequent inductively coupled plasma (ICP) etching of Cr that utilizes Cl$_2$:O$_2$ plasma with gas proportions of 80:20 sccm, and operates with an ICP power of 500 W and at a chamber pressure of 5 mTorr. Next, ICP etching of SiO$_2$ with Cr hard mask is performed with CHF$_3$:Ar plasma with gas proportions of 40:10 sccm, and operates with an ICP power of 800 W and at a chamber pressure of 10 mTorr. Note that the ICP etching of SiO$_2$ is crucial in obtaining smooth and straight grating sidewall of the SiO$_2$ stamp. Lastly, Cr etchant (CR-7S) is used to remove the Cr mask.

To prepare for NIL, the SiO$_2$ stamp is first coated with an anti-adhesion monolayer of perfluorodecyltrichlorosilane (FDTS) in n-heptane solvent for 5 min, which prevents perovskite from sticking to the stamp after NIL. Next, the stamp is rinsed with acetone for 1 min and blow dried with N$_2$, and then annealed on a hotplate for 20 mins at 100 °C to enhance the adhesion of FDTS layer on the stamp, remove moisture in FDTS and increase the hydrophobicity of FDTS. After that, the stamp is placed on the perovskite thin film coated substrate. The NIL process, performed with Obducat nanoimprinter, utilizes a multi-step process. During the 500 s heat-up time, the pressure is increased from 10 bar to 70 bar at an interval of 10 bar, while the temperature is increased from 35 °C to 100 °C at an interval of ~75 s. An imprint time of 20 mins is then employed at 100 °C and 70 bar, after which the system is cooled down and the pressure is released slowly. The NIL process is then finished, and perovskite nanostructures are formed as a negative replication of the stamp. The details of the fabrication process is shown in a step-by-step manner in Supplementary Figures 1 and 2.

**Steady-state micro-photoluminescence (micro-PL) measurement**

To characterize the perovskite lasers, micro-photoluminescence measurement (with setup schematically shown in Supplementary Figure 5) is carried out at room temperature, in ambient atmosphere without any thermal control. A 355 nm Nd:YV04 laser (Spectra-Physics Talon-355-20) under CW operation is used as the pump source. To rotate the predominantly vertically polarized output of the laser (100:1 vertical) to horizontal polarization, a half-wave plate with its axis 45 degrees with respect to the vertical direction is used. The incident pump power intensity on the sample is adjusted using a variable neutral density (ND) filter and monitored with a thermal power sensor (Ophir 7Z01250). A longpass dichroic mirror (50% transmission/reflection at 425 nm) is employed to selectively reflect the pump light while allowing the longer-wavelength emitted light to pass through. The horizontally polarized pump beam is delivered to the device under test through a ultra-violet (UV) microscope objective with a numerical aperture (NA) of 0.13 and the dichroic mirror. In turn, this objective also serves to collect the emitted radiation. To minimize chromatic aberration, a telescope is introduced in the pump path for beam-shaping, such that the focal planes of the pump and emission wavelengths coincide. Using a cascaded 4-f imaging system in conjunction with a pump filter, laser structures are either imaged onto a visible CCD camera (Ophir SP90281), or spectroscopically measured with a spectrograph (Princeton Instruments IsoPlane SCT-320) coupled to a cooled Si detector (Princeton Instruments PIXIS:400BRX). The spectra used in constructing the spectra evolution, light-light curve and emission wavelength evolution (Fig. 3) are taken with a wavelength step-size of 0.19 nm. Note that although the perovskite DFB cavities have pattern size of 100 μm x 100 μm as defined by the pattern size of the stamp, only 20 μm x 20 μm patterned area is pumped in emission characterizations, therefore reducing the effective device footprint.